\pdfoutput=1

\documentclass[aps,showpacs,preprintnumbers,amsmath,amssymb,
eqsecnum, twocolumn, tightenlines
]{revtex4}

\usepackage{graphicx}

\newcommand{\be}{\begin{eqnarray}}
\newcommand{\ee}{\end{eqnarray}}

 \newcommand{\gsim}{\mathrel{\hbox{\rlap{\lower.55ex \hbox {$\sim$}}
                   \kern-.3em \raise.4ex \hbox{$>$}}}}
\newcommand{\lsim}{\mathrel{\hbox{\rlap{\lower.55ex \hbox {$\sim$}}
                   \kern-.3em \raise.4ex \hbox{$<$}}}}

\def\roughly#1{\mathrel{\raise.3ex\hbox{$#1$\kern-.75em%
\lower1ex\hbox{$\sim$}}}}
\def\lsim{\roughly<}
\def\gsim{\roughly>}

\newcommand{\ba}{\begin{eqnarray}}
\newcommand{\ea}{\end{eqnarray}}

\begin{document}


\title{ Jet Quenching via Gravitational Radiation in Thermal AdS }
\author { Edward Shuryak, Ho-Ung Yee, and Ismail Zahed }
\address { Department of Physics and Astronomy, State University of New York,
Stony Brook, NY 11794}
\date{\today}

\begin{abstract}
We argue that classical bulk gravitational radiation effects in AdS/CFT, previously ignored because of their subleading nature
in the $1/N_c$-expansion, are magnified by powers of large 
Lorentz factors $\gamma$ for ultrarelativistic jets, thereby dominating other forms of jet energy loss in holography at finite 
temperature. We make use of the induced gravitational self-force in thermal AdS$_5$ 
to estimate its effects.  In a thermal medium,  relativistic jets may loose most of their energy 
through longitudinal drag caused by the energy accumulated in their nearby field
as they zip through the strongly coupled plasma.

\end{abstract}
\maketitle

\section{Introduction}

This paper stems from two different motivations.
One is a (century old) quest to understand the radiation reaction force
in classical electrodynamics and general relativity, especially in higher space-time dimensions. The second (practical) motivation is the need to estimate the magnitude of such
a classical force on a falling gravitational body in thermal AdS$_5$ space-time. This latter problem is related to the empirically
important issue of jet quenching in strongly coupled quark-gluon plasma as unravelled by recent collider experiments 
at RHIC and LHC. 

Because of these two different motivations, we make the introduction a bit large to allow for a brief descriptive
of the history of both subjects which is perhaps needed for some diverse readers. A more thorough analysis of the
radiation reaction force in electrodynamics in higher even-space+time dimensions will be presented in 
a companion paper~\cite{selfforce}.

\subsection{Radiation reaction force}

One general theoretical problem, with which theorists such as Abraham, Lorentz, Dirac and others
struggled from about the beginning of 20-th century, deals with the understanding of the radiation 
reaction force.  The concept of non-relativistic and relativistic classical radiation loss can be found 
in most textbooks on classical electrodynamics. For instance, the relativistic form of the Larmor 
formulae for the power radiated by an accelerated charge reads
\be 
P=-{2e^2 \over 3}  \, \ddot{x}\cdot  \ddot{x},
\label{eqn_Larmor} 
\ee 
with  $x^a(\tau)$ being the particle trajectory parameterized by its proper time $\tau$. The double dots 
reflect the proper accelaration. It is not difficult to write down a longitudinal force that drags the
charged particle matching the energy loss (\ref{eqn_Larmor}). This procedure refers to the radiation
or large distance derivation of the local foce, based essentially on the energy-momentum flux 
radiated on an asymptotic surface at infinity (large sphere). However, there are many reasons for why a local 
derivation  of the self-force that only makes use of the local fields acting on the 
accelarated charge is warranted.

There is a large body of research on this approach for both scalar, vector (electromagnetism) 
and tensor (gravity) radiation both in flat and curved space-time. Once the energy loss is accounted
for, the radiation reaction force is fixed by the condition of balancing loss. Putting aside some known
paradoxes related to this particular construction, we would like to focus on the following specific questions:\\\\
(i)  Is the radiation reaction force amenable to a local definition from a regulated self-field at
the instantaneous position of the charge?\\ (ii) In electrodynamics the radiation reaction force is expressed via the acceleration, but
in general relativity particles move along geodesics for which the covariant acceleration is zero. 
So what makes up the force?\\ (iii) The Larmor
formula involves the acceleration thanks to the dipole approximation in non-relativistic electrodynamics,
but what about higher derivatives in higher dimensional settings and general relativity?

While we do not address all of these difficult issues here, we aim at exploring the practical
importance of some theoretical proposals already in the literature. If (i) is true, it
would simplify calculations considerably, especially in the case of general relativity since
defining distant surfaces in curved backgrounds is in general elusive. More specifically, can the 
radiation reaction force be the {\it self-interaction force} following from the charge's own field 
due to its past trajectory?

 The closest to formulating an answer to a local definition of the self-force in four-dimensional
 gravity was the work of Mino, Sasaki and Tanaka and also Queen and Wald
 ~\cite{Mino:1996nk,Quinn:1996am}, where the gravitational
 radiation reaction force was obtained in a closed form
 \be
&&m{\ddot{x}}^a=Gm^2\dot{x}^b\dot{x}^c\,\int_{-\infty}^{\tau_-}\,d\tau'\,\label{XG1}\\
&&\left(\frac 12\nabla^a{\bf G}^-_{bca'b'}-\nabla_b{\bf G}^{-a}_{c\,\,\,\,a'b'}-\frac 12
\dot{x}^a\dot{x}^d\nabla_d\,{\bf G}^-_{bca'b'}\right)\,\dot{x'}^a\dot{x'}^{b},
\nonumber
\ee
whereby the integral is carried over the past world line of the particle up to a regulated time $\tau_-$. Here
${\bf G}^-$ is the retarded propagator for the Einsten equation with the particle's own stress tensor acting as a source. While
more details on this equation will be given below, here we just note that the bracket in (\ref{XG1}) is just
the Christoffel's force from a metric perturbation as induced by the past history of the particle. Note that the result (\ref{XG1}) was derived
in harmonic gravitational gauge: in this work we will ignore all subtleties that may be associated with this particular gauge choice.

Does this force or its simplified analogues really work? Clearly, for cyclotron radiation in flat four dimensional
space-time,  a force given by an analogous expression would vanish. The simplest way to see this is to note that the retarded propagator in flat four-dimensional space-time reads
\be
{\bf G}^-(x,x')=\frac 1{2\pi}\,\theta (x^0-{x'}^0)\,\delta((x-x')^2),\nonumber
\ee
which is totally localized on the light cone. As a result, there are simply no points on the cyclotron world line
that intersect the past light cone.  However this  drawback is short lived if were to notice: 1) In curved space-times the retarded propagator gets contributions from light rays reflecting on the curved background thereby causing intersections
with the past light cone; 2) In even-space+time dimensions the retarded propagator has support on the entire interior of
past light cone even in flat-space, causing contributions from the past history of the particle.
(\ref{XG1}) is a valid local radiative self-force in gravity.

We will use (\ref{XG1}) to address the issue of jet
quenching in AdS/CFT correspondence. Our main observation is that for ultrarelativistic jets, (\ref{XG1}) is to be used in the regime 
of very short proper times such that the Taylor expansion of all the factors in the numerator in proper time would provide the
first finite term after all infinities are subtracted and renormalized.  In our analysis, the magnitude of the proper time  $\epsilon$ is
\be 
\epsilon \approx 1/\gamma^2 \ll 1, 
\label{XG2}
\ee
as will be made clear in later sections. In the small $\epsilon$-expansion the radiative
self-force in curved and thermal AdS$_5$ follows from the finite term in (\ref{XG1}) after using the 
leading singularity of the retarded propagator ${\bf G}^-$ with the appropriate Taylor coefficient.

\subsection{Jet quenching}
 
  Jets are the most spectacular aspect of  QCD physics in high energy setting. Their interaction with the
  quark-gluon plasma in heavy ion collision has long been deemed very interesting, starting from the 
   early work by Bjorken~\cite{Bjorken:1982tu}.  Indeed the phenomenon known as jet quenching
  has been observed  in heavy ion collisions at RHIC; see early summary in Refs.\cite{Adams:2005dq,Adcox:2004mh}. The observed magnitude of jet quenching was significantly stronger than predicted by the perturbative estimates.
  
  More spectacular jet quenching phenomenon has been reported in the first LHC heavy ion run
  in November 2010: events with huge energy ($\approx 100$ GeV) having been lost and 
  dissipated into the quark gluon plasma. Many surviving jets retaining a fraction of the initial energy  
  and travelling back-to-back to the trigger jet appear to fragment outside the medium in the standard way. 
  There is no indication of collinear radiation in the jet cone or nearby. The energy loss appears to 
   be due to some longitudinal force applied to the moving charge as opposed to the expected transverse
   kicks due to perturbative QCD (pQCD). 
   
   These observations strongly suggest the notion that the quark-gluon plasma is strongly coupled and that
    it needs to be treated in the AdS/CFT setting in trying to understand the mechanism of energy loss. Some of the previous works on this line will 
    be briefly reviewed below. What seems particularly important is the dependence of the jet quenching
    on the jet energy $E$. For discussion purposes, let us assume that the energy loss is of some power $p$ of the energy: 
\be 
dE/dx \approx  E^p.
\ee
In pQCD the energy loss through radiation results in splitting of the leading parton into many partons, each carrying a
fraction of the leading energy. The scaling nature of pQCD implies that jet quenching is proportional to the energy itself;  
$-dE/dx \approx  E$ with $p=1$. 
This is radically different from e.g.
%
%
 standard synchrotron radiation in accelerators which has a power  $p=4$. As noted by Pomeranchuck 
 in 1939 and more recently by Kharzeev \cite{Kharzeev:2008he}, the energy loss with $p>1$ leads to the following 
  \be 
   {1 \over (p-1)E_f^{p-1}}= {1 \over (p-1)E_i^{p-1}}+\int k(x) dx .
   \ee
The interesting result is a $finite$ final energy $E_f$ even if the initial jet energy is very large $E_i\rightarrow\infty$. If so, then all
  companion jets will  have nearly same energy, independent of the energy of the original (trigger) jet!
  
As we will show in this paper, the self-force from classical gravitational radiation of a falling ultrarelativistic
object in the holographic bulk does in fact produce a power $p>1$  of the Lorentz factor. Thus this interesting behavior with higher energy jets stopping 
at shorter distances seems to be also a feature of the AdS/CFT correspondence.

\subsection{Jet quenching in AdS/CFT } \label{sec_jet_ads}

The idea that strong coupling physics is at work in jet quenching and could be addressed in the context of 
AdS/CFT was initially put forward by one of us (with Sin)~\cite{Sin:2004yx}. Since then, the number of
applications and discussions along these lines is legion. For completeness, we briefly review these discussions 
of jets in the AdS/CFT context.

In QCD the simplest process in which jets have been first seen is $e^+e^- \rightarrow $ hadrons. The newborn quark-antiquark pair
move in random direction, and create two back-to-back jets. Its description started from string breaking models at low energies 
(e.g. the so called Lund model) and later developed into an elaborate pQCD description of the partonic coherent cascade at high energies. The outcome is a good description of the data.   

Therefore it is not surprising that the first papers addressing some of the issue in AdS/CFT set-up  have the same setting: two charged particles moving with velocity $v$ back-to-back from each other~\cite{Lin:2006rf, Lin:2007pv}.
(The same problem with non back-to-back velocities was addressed earlier in the
context of the pomeron problem~\cite{Rho:1999jm}). 
As the charged particles recede in thermal AdS,  a light-like string between them falls gravitationally in the bulk. The
gravitational hologram on the boundary CFT
shows no sign of a falling string, but rather a point-like explosion. Indeed, the boundary
energy-momentum tensor $T^{\mu\nu}$ in the comoving frame yields a spatial (pressure) part which is not isotropic. The falling
string has nothing to do with a hydrodynamical explosion as no black-hole is formed. The strong coupling regime does not tolerate jets, even if the two charges move by straight lines as in  $e^+e^- \rightarrow $ hadrons. More details about this point-like explosion, termed as conformal collider physics, was discussed by Hofman and Maldacena \cite{Hofman:2008ar}.

Ultrarelativistic jets in holography are well approximated by light-like rays travelling near the UV 
boundary of AdS. Their demise comes at the hands of holographic bulk gravity which cause them
to ultimately fall towards the horizon and be absorbed by the standing black hole.
This scenario was put forward originally in~\cite{Sin:2004yx}, and recently revisited and sharpened by 
many~\cite{Gubser:2008as,Chesler:2008uy,Arnold:2011qi} (and references therein).   Although a number of authors have considered gluons or quarks as 
end-of-the-string objects~\cite{Gubser:2008as,Chesler:2008uy} falling on a static thermal black hole
in AdS, it turns out that to assess the jet penetration length it is simpler to approximate the end point by
a path of near-null geodesic as initially suggested in~\cite{Sin:2004yx}. 

The current issue on jet physics in the context of the AdS/CFT set up has now shifted to the question of 
what exactly are the masses of these objects and how the jet process is initiated.
One of the main results is that the shape of the geodesic describes ``scale evolution" of a jet, and is related to 
the jet quenching strength versus the length $x$ travelled since its production. The geodesic path 
consists of two parts: a near horizontal one suggestive of a very nonlinear rise of energy loss $-dE/dx \approx x^2$, 
followed by a near-vertical plunge into the horizon, finished by a jet explosion as it splashes on the black hole horizon. 
Needless to say, both results need to be carefully tested against the phenomenologically accessible
information at both RHIC and LHC. We will postpone this to the future.

\subsection{Cyclotron versus gravitational radiation}

The first step toward relating two very different motivations mentioned in the earlier part of introduction has been done by one of us (with Khriplovich) nearly 40 years ago \cite{Khriplovich:1973qf}, applying the same method to 4 problems: cyclotron electromagnetic/gravitational radiations in flat or curved 3+1 dimensional spaces in the ultrarelativistic regime $\gamma \gg 1$. The results for the radiation intensity are
  \begin{eqnarray}
 &&{\bf I}_{\rm e.m.}^{\rm flat} \approx e^2 \gamma^4/R^2,     \qquad {\bf I}_{\rm grav}^{\rm flat} \approx G_4 m^2\gamma^4 /R^2, \nonumber\\ 
  &&{\bf I}_{\rm e.m.}^{\rm curv} \approx  e^2 \gamma^2/R^2,\qquad      {\bf I}_{\rm grav}^{\rm curv} \approx  G_4 m^2\gamma^2 /R^2,
  \label{COMP}
  \end{eqnarray}
 where the Schwartzschield metric was used for the curved space. An ultrarelativistic particle was set to rotate at the (unstable) circular orbit of radius $R\approx (3/2)r_h$.  Note the decrease of the powers of $\gamma$ as we go from cyclotron radiation in flat space to gravitational radiation around a Schwartzchield black hole. Also note that the powers of energy in the formulae consistently exceed one. 
 In the last case of (\ref{COMP}), Newton's constant $G_4$ appears in combination with the particle energy 
 $E=\gamma \,m$. As a result, if the particle is small and the massless limit $m\rightarrow 0,\gamma\rightarrow\infty, m\gamma=$drag
 fixed is considered, the limit is non-singular and thus the answer is the only possible one which makes sense physically.
  
For pedagogical reasons, we would like to discuss the problem, i.e. gravitational radiation and self-force in {\em $AdS_5$ }, in two steps. In this subsection we will compare the settings of the cyclotron radiation with that in pure or empty AdS$_5$.
Then in later section, we will introduce thermal $AdS_5$, and study our problem in two different coordinate frames complementarily; the plasma rest frame and the jet comoving frame.

The familiar setting for cyclotron radiation consists of a point charge rotating on a circle of fixed radius $R$ with
$\gamma=E/m\gg 1$.  The ratio of the proper time $\tau$ to the coordinate time $t$ is fixed, and e.g.  
\be  
\dot{x}^a={dx^a\over d \tau}=\gamma {dx^a\over dt}. 
\ee 
The 4-velocity is tangent to the world line, while the accelaration (induced by appropriate magnetic field) is orthogonal to it and is directed radially toward the
circle's center. All odd derivatives of coordinates share same direction with $ \dot{x}^a$ while all even ones with $ \ddot{x}^a$. An extra power of gamma appears for each derivative
over the proper time. While the  cyclotron radiation is a textbook example in 3+1 dimensional space-time, it is not so in other dimensions.  The difference, in respect to the self-force issue, will be discussed in 2+1 and 4+1 space-times in the companion paper \cite{selfforce}. We note that these space-time dimensions are the relevant ones for holographic models of QCD in 1+1 and 3+1 dimensions.

 \subsection{Motion in Schwartzschield versus AdS$_5$ space-times}

In the gravity setting the problem discussed in Ref.\cite{Khriplovich:1973qf}
is that of an ultrarelativistic particle rotating in (unstable) circular orbit of a 3+1 dimensional Schwartzschield metric 
in polar coordinates $(t, r, \theta, \phi)$.  In contrast, an empty AdS$_5$ space is defined by the metric
\be 
ds^2= \frac 1{z^2}\left[ -dt^2 + (dx^1)^2 + (dx^2)^2+ (dx^3)^2+dz^2  \right],
\ee
which is a near-horizon limit of the 9+1 dimensional black brane solution. The brane itself is extended in 3 coordinates 
$x^1,x^2,x^3$, and the metric is independent of them. These three coordinates are the analogue
of the polar coordinates $\theta,\phi$ of the Schwartzschield metric, with $z$ playing the role 
of the radial coordinate $r$. Unlike the Schwartzschield black hole, the empty $AdS_5$ has no horizon since it originates from the
so called extremal black hole with maximal possible charge.

In the next section we will derive an expression for the self-force containing multiple derivatives of the local 
trajectory over the proper time based on (\ref{XG1}). To understand its origin, we recall the geodesic equation
for free falling in general gravity
\be 
\ddot{x^a}=-\frac 12 \,g^{ab}\, ( g_{cb,d}+ g_{db,a}- g_{cd,b})\, \dot{x^c} \dot{x^d}, 
\ee
where the right-hand side is driven by the Christoffel symbols.
Since the Schwartzschield metric  depends only on $r$, and the empty $AdS_5$ metric only on $z$, the acceleration
is directed only along these coordinates. The large tangent components of the 4-velocity on the right-hand side are of
order $\gamma^2$ and so is the accelaration.

For the Schwartzschield case, equatorial geodesics follow by setting $\dot{x}^\theta=u^\theta=0$. Thus
\be  
\ddot{x^r} = - \frac 12 \left(1-\frac {r_h}r \right) \, \left( -{r_h \over r^2} \,(u^0)^2 + 2r (u^\phi)^2\right),
\ee
using $\dot{x}=u$.  The radius of interest for the ultrarelativistic circular orbit is a textbook result
\be 
r_c=\frac {3r_h}2 \left(1+{1\over 2 \gamma^2}\right).
\ee
Repeated differentiations of $r_c$ yields further derivatives over proper time $\tau$ if needed. Analogous expressions for
AdS$_5$ can readily be obtained using similar arguments after setting the path in the geodesic inside the $(t,x^1,z)$ plane
with $\dot{x}^2=\dot{x}^3=0$.  

 In order to find the geodesic path itself, it is however more convenient to use existing integrals of motion
 rather than solving the second order differential equations for the acceleration just mentioned. The Schwartzschield metric
does not depend on $t, \theta,\phi$ and thus the lower-indexed momenta (per mass) $u_0,u_\theta,u_\phi$ are conserved:
those can be called the energy and two components of the angular momentum, respectively. The $AdS_5$ metric is independent of the 4 coordinates of the (boundary Minkowski world) $(t,x^1,x^2,x^3)$. As a result, all 4-momentum components $u_a=P_a,a=1,2,3,4$ are conserved.

 The geodesic line follows from the on-shell condition for the squared 4-velocity 
\be 
g_{00}(u^0)^2+ g_{rr} \dot{r}^2 + g_{\phi\phi}(u^\phi)^2=RHS,
\ee
where $RHS=-1$ for a massive particle and 0 for a massless one, and expressing all components in terms of the conserved 
momenta. For the Schwartzschield case one should use $u^0=g^{00} \gamma, u^\phi= g^{\phi\phi} L$ and for $AdS_5$ 
\be u^a=g^{aa} P_a, \,\,\, a=1,2,3,4. 
\ee
As a result, the respective radial velocities $\dot{r}$ or $\dot{z}$ can be expressed in terms of the coordinates. 
The result is a first order equation that is readily integrated. For instance, the trajectory in AdS$_5$ is
\be  
{dx^5\over dx^1}={\dot{z}\over \dot{x}}= \frac 1P \sqrt{(E^2-P^2)-1/z^2},
\ee 
which integrates to
the path
\be 
x^1=\int_{z_i}^z\,dz'\,\,\frac{P}{\sqrt{E^2-P^2-1/{z'}^2}}. 
\ee
There is a qualitative difference between falling in a Schwartzschield background and in an AdS$_5$ background. Indeed, while
in the former a trapped circular orbit is possible, it is not in the latter. Particles fall indefinitely in AdS$_5$ irrespective of their
initial conditions. 
A jet in AdS$_5$ assumes a particle
being very energetic with $\gamma\gg 1$ along $x^1$ with 4-momentum conservation.  The falling trajectory is composed of
two parts: 1) an almost straight initial path with the x-momentum much larger than the z-momentum with $\dot{x}_1 \gg \dot{z}$;
2) a final diving or almost vertical plunge into the horizon. 

Falling in the Schwartzschield and in the AdS$_5$ involves transverse spatial acceleration which is normal to the velocities 
(at least initially in AdS$_5$). In the Schwartzschield case the circular fall produces a radiation intensity of order $\gamma^2$ as shown in (\ref{COMP}). 
Is the fall in the empty AdS$_5$ also accompanied by a $\gamma^2$ radiation? The answer is negative and one can readily see that there is no dependence of the radiation on $\gamma$.  Indeed, since the AdS$_5$ metric 
is Lorentz invariant, we can switch to a frame where $P_1=0$. The trajectory is then confined to the $(t,z)$ plane with no 
transverse acceleration. This is not the case for thermal AdS$_5$ as we discuss in later sections.

\section{Self-force in general relativity}

The local self-force in 3+1 gravity with zero cosmological constant was derived originally
by Mino, Sasaki and Tanaka and also Queen and Wald~\cite{Mino:1996nk,Quinn:1996am}.
As we noted in the introduction and now we repeat for completeness,
\be
&&m{\ddot{x}}^a=G_5 m^2\dot{x}^b\dot{x}^c\,\int_{-\infty}^{\tau_-}\,d\tau'\,\label{G1}\\
&&\left(\frac 12\nabla^a{\bf G}^-_{bca'b'}-\nabla_b{\bf G}^{-a}_{c\,\,\,\,a'b'}-\frac 12
\dot{x}^a\dot{x}^d\nabla_d\,{\bf G}^-_{bca'b'}\right)\,\dot{x'}^a\dot{x'}^{b},
\nonumber
\ee
with ${\bf G}^-$ being the graviton retarded propagator,
\be
\Box  {\bf G}_{aba'b'}^- -2\,{{\bf R}^c}_{ab}^{\,\,\,\,\,\,d}\,{\bf G}_{cda'b'}^-=
-16\pi\,\overline{g}_{aa'}\overline{g}_{bb'}\,\delta_5(x,x'),\nonumber\\
\label{G1X}
\ee
where $x=x(\tau)$, $x'=x(\tau')$,  $\delta_5(x,x')=\delta^5(x-x')\sqrt{-g}$
and $\overline{g}$ is DeWitt's bilocal for
parallel displacement along the geodesic~\cite{DEWITT}. 
Although the original derivation of (\ref{G1}) was carried in 3+1 dimensional space with
zero cosmological constant and matter, its physical interpretation is applicable in any dimensions: the right-hand side is  simply a modification of Christoffel symbols due to the retarded metric perturbation of the particle trajectory. Therefore we assume it to hold in general, especially for thermal AdS$_5$  in 4+1 dimensions.

For ultrarelativistic jets, the eikonal limit is appropriate
\be
{\bf G}^-_{aba'b'}(x,x')\approx  16\pi \,\overline{g}_{aa'}\,\overline{g}_{bb'}\,{\bf G}^-(x,x'),
\label{G2}
\ee
with $\Box{\bf G}^-(x,x')=-\delta_5(x,x')$. Inserting (\ref{G2}) in (\ref{G1}) yields
\be
&&m\dot{\dot{x}}^a\approx 4\pi^2\,G_5\,m^2\,\int_{-\infty}^{\tau_-}\,d\tau'\,\label{G3}\\
&&\left((\nabla^a{\bf G}^--\dot{x}^a\dot{x}^d\nabla_d{\bf G}^-)(\dot{x}\cdot\dot{x'})^2
-2\,\dot{x}\cdot\dot{x}'\dot{x'}^a\dot{x}^d\nabla_d{\bf G}^-\right).
\nonumber
\ee
In the above, we have dropped terms of the type $\nabla^a\overline{g}$ as they are subleading in 
small proper time $\epsilon$-expansion than $\nabla^a{\bf G}^-$. 

The scalar retarded propagator in a curved background of 4+1 dimensions can be related to the one 
in 2+1 dimensions. Explicitly,
\be
{\bf G}^-(x,x')=-\frac 1{2\pi}
\frac{d}{d\sigma}\left(\Theta (x',x) \frac {\theta (-\sigma)}{2\pi}\frac{\sqrt\Delta}{\sqrt{-2\sigma}}\right),
\label{G4}
\ee
where the expression inside the bracket is the retarded propagator in 2+1 dimensions. 
$\Theta(x',x)$ is the generalized heaviside step-function with
a space-like surface through $x$ (the final form (\ref{G4}) doesn't depend on the choice of this surface), and $\Delta$ is the Van-Vleck determinant 
\be
\Delta (x,x')= \left(g(x)g(x')\right)^{1/2}\,{\rm det}\left(\nabla_a\nabla_{a'}\sigma (x,x')\right),
\label{G6}
\ee
which is a scalar two-point function.  $\sigma$ is an another two-point
scalar function 
\be
\sigma(x,x')=\frac 12 (\tau-\tau')\int_{\tau'}^\tau\,d\tau''\,\dot{x}(\tau'')\cdot\dot{x}(\tau''),
\label{G5}
\ee
which is defined by the geodesic between $x$ and $x'$. It is negative for time-like geodesics
\be
\sigma(x,x')=-\frac 12(\tau-\tau')^2=\frac 12 {d(x,x')^2},
\label{G5X}
\ee
where $d(x,x')$ is the chordal distance. While the latter is only defined locally
for general curved space-times, it can be defined globally for dS and AdS spaces because
of their spherical and hyperbolic nature. Indeed, for AdS$_5$ the finite distance is
\be
{\rm cos}\,(d(x,x'))-1=\frac{({\bf x}-{\bf x}')^2}{2zz'},
\label{GXX}
\ee
with z being the conformal direction and ${\bf x}^2=-t^2+x^2$.  This relation is readily 
derived by embedding AdS$_5$ in $R^6$ with a hyperbolic constraint.
We further note that the retarded propagator in  AdS$_5$ is the known function of (\ref{G4}) 
 derived in~\cite{Danielsson:1998wt}, and our generic small time expansion is consistent 
 with it.

\section{Self-force on Jets in AdS$_5$}

\subsection{The expansion in $\epsilon$}
For ultrarelativistic jets, the trajectory is characterized by small proper times, and one can expand in it.
Inserting (\ref{G4}) into (\ref{G3})  and following our arguments for the self-force in 4+1 dimensions show
that the gravitational self-force for ultrarelativistic jets is dominated by the leading singularity in the (covariant) gradient
of the propagator, 
\be
\nabla^a{\bf G}^-\approx -\frac{3}{4\pi^2}\,\frac {\sqrt{\Delta}\,\sigma^a}{\epsilon^5},
\label{G7}
\ee
with small $\epsilon=(\tau-\tau')\ll 1$.  The smallness of $\epsilon$ will be explained further below.
The problem is then reduced to a covariant expansion of  $\sqrt{\Delta}$ for $x'$ near $x$.
For that we follow~\cite{DEWITT,Christensen:1976vb} and expand $\sqrt{\Delta}$ first covariantly in terms of 
the world function $\sigma$ and then proceed to Taylor expand $\sigma$. 
Specifically
\be
&&\sqrt{\Delta}=1+\frac 1{12}\,{\bf R}_{ab}\,\sigma^a\sigma^b-\frac 1{24}\,{\bf R}_{ab; c}\,\sigma^{a}\sigma^b\sigma^c\\
&&+\left(\frac 1{288}\,{\bf R}_{ab}{\bf R}_{cd}+\frac 1{360}\,{{{{\bf R}^m}_{a}}^{n}}_{b}{\bf R}_{mcnd}
+\frac 1{60}\,{\bf R}_{ab; cd}\right)\nonumber\\
&&\times \sigma^a\sigma^b\sigma^c\sigma^d+\cdots,\nonumber
\ee

For both empty and thermal AdS spaces there are significant simplifications, which come from the fact that the Ricci tensor
${\bf R}_{ab}$ is proportional to the metric, since the Einstein equation with cosmological constant tells us
\be
{\bf R}_{ab}-\frac 12 g_{ab}\,{\bf R}=\frac {\Lambda}{2}\,g_{ab},
\ee
with the scalar curvature ${\bf R}=-5\Lambda/3$. As a result, 
the covariant derivatives of the Ricci tensor simply vanish
\be 
{\bf R}_{ab;c}={\bf R}_{ab;cd}=0, 
\ee
which yields
\be
\sqrt{\Delta}=&&1-\frac 1{18}\sigma+\frac 1{648}\sigma^2
\label{Bp1}\\&&+\frac 1{360}\,{{{{\bf R}^m}_{a}}^{n}}_{b}{\bf R}_{mcnd}
\,\sigma^a\sigma^b\sigma^c\sigma^d+\cdots,\nonumber
\ee
after using the identity $\sigma^a\sigma_a=2\sigma$. 

For a falling jet in AdS$_5$ a tremendous simplification takes place when rewriting (\ref{Bp1}) 
in terms of jet trajectory proper time difference. Indeed, since the 
falling jet follows a geodesic at zero'th order, the bilocal $\sigma(x,x')$ uniquely determines the 
distance crossed by the jet for short or small $\epsilon$. The geodesic definition
and length are unique when $x'$ is in the neighborhood of $x$. Thus $\sigma=\epsilon^2/2$ and 
$\sigma^a=\epsilon\,\dot{x}^a$ where $\dot{x}^a$ is tangent to the jet geodesic.

Another way to say this is to note that the covariant Taylor expansion of $\sigma$
along the particle trajectory (and not the graviton trajectory) to fourth order is 
\be
\sigma\approx -\frac{\epsilon^2}2-\frac {\epsilon^4}{12}\,D\dot{x}\cdot D\dot{x},
\label{G8}
\ee
with $D\dot{x}^a\equiv \ddot{x}^a+\Gamma^{a}_{bc}\dot{x}^b\dot{x}^c$,
the covariant or long derivative. The covariant derivative  and its higher order variants vanish along the
falling jet geodesic making the first contribution exact.  Thus along
geodesic jet trajectories in AdS$_5$
\be
\sqrt{\Delta}\approx 1
+\frac{\epsilon^4}{360}\,{{{{\bf R}^m}_{a}}^{n}}_{b}{\bf R}_{mcnd}
\dot{x}^a\dot{x}^b\dot{x}^c\dot{x}^d,
\label{G10}
\ee
neglecting renormalizations taking care of divergences and subleading terms in the large $\gamma$-expansion.
In terms of (\ref{G10}) the gravitational self-force (\ref{G3}) simplifies
\begin{eqnarray}
m{\ddot{x}}^a\approx-\frac{ G_5 m^2}{30\pi}\,\left(\int d\epsilon\right)
\,{{{{\bf R}^m}_{e}}^{n}}_{b}{\bf R}_{mcnd}\,\,
\dot{x}^e\dot{x}^b\dot{x}^c\dot{x}^d\,\,\dot{x}^a,\nonumber\\
\label{SELF}
\end{eqnarray}
The final integration along the proper time $\int d\epsilon$ will be explained shortly, and is of order $1/\gamma^2$.
The dragging gravitational covariant self-force on the falling jet is longidudinal and of order
$\gamma^3$. It is entirely driven by a (squared) Riemann tensor.

\subsection{Final integration and magnitude of $\epsilon$ }

So far we have assumed $\epsilon$, the proper time between the emission of the graviton and its action on the charge,
to be small. We have not yet specified why and how small it is. A qualitative answer follows from the fact that the more
relativistic the jet, the more localized its own trailing field. Indeed, the localization in time is expected to be 
$t\approx 1/\gamma$  and thus a localization in propertime $\tau\approx \epsilon\approx 1/\gamma^2$.

A way to see this is to note that the final $\epsilon$ integration needs to be cutoff at {\rm large} $\epsilon$.  
Indeed, (\ref{G8}) in flat space-time reduces to

\be
\sigma|_{\rm flat}\approx -\frac{\epsilon^2}2-\frac {\epsilon^4}{12}\,\ddot{x}\cdot \ddot{x}
\label{G8X}
\ee
For ultrarelativistic motion, we can make the substitution

\be
\epsilon\rightarrow \epsilon\left(1+\frac 1{6}\,\epsilon^2\,\ddot{x}\cdot\ddot{x}\right)^{1/2}
\label{IN}
\ee
leading to a finite resummed result

\be
\int_0^\infty\,d\epsilon\,
\left(1+\frac 1{6}\epsilon^2\,\ddot{x}\cdot\ddot{x}\right)^{-5/2}=\frac 4{\sqrt{6\,\ddot{x}\cdot\ddot{x}}}
\label{COEF}
\ee
with most of the contribution stemming from the range $\epsilon \approx 1/\gamma^2$.  As a result, the dropped
terms in our quasi-local analysis of the gravitational self-force are all subleading in $1/\gamma$. This argument
shows how a schematic resummation of the subleading corrections  yields a finite result for $\int d\epsilon$.  The
qualitative character of this subtitution does not fix the overall coefficient exactly. For that, more quantitative
work is needed.

The scaling of $\epsilon\approx 1/\gamma^2$ is also in effect 
in the synchrotron analysis in even-space+time dimensions as we discuss in the
companion paper~\cite{selfforce}. This underlines again the analogy between cyclotron and gravitational radiation
as we have noted above.

 \begin{figure}[!t]
\begin{center}
\includegraphics[width=7cm]{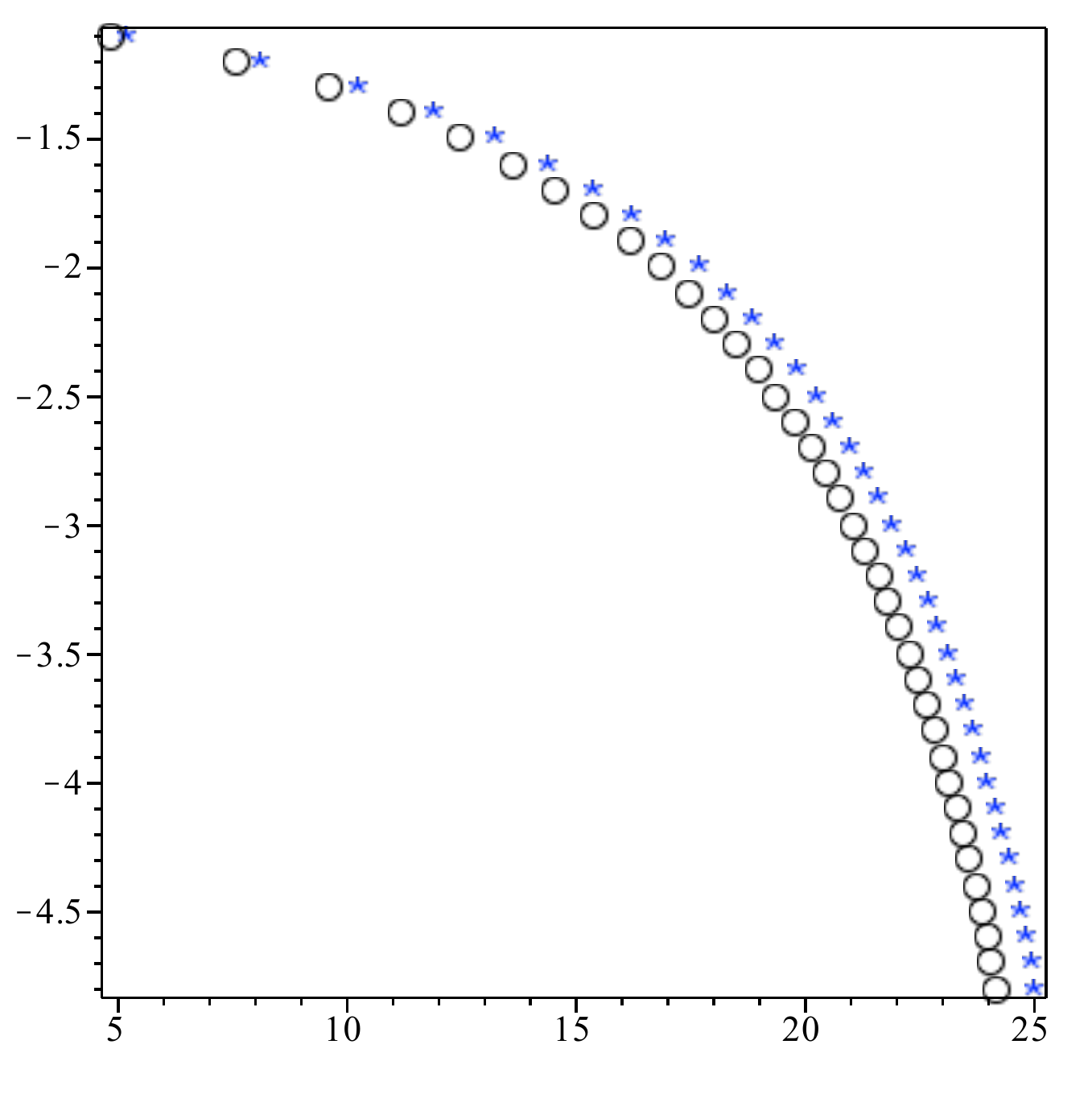}
\end{center}
\vspace{-5ex}\caption{(color online) Example of a geodesic trajectory for a falling particle in the (inverted) 5-th coordinate $-z$ as a function of $x^1$. The open (black) circles are for massive particle, (blue) asteriks are for massless geodesic. The parameters are explained in the text
}\label{fig_2geo}
\end{figure}

\subsection{Holographic gravitational force estimate}

The previous analysis is purely 5-dimensional, and to apply that to AdS/CFT, we have to translate 5-dimensional results
into 4-dimensional gauge theory ones.
The holographic duality holds in the double limit in which the number of colors is large compared to 't Hooft coupling $N_c\gg \lambda=g^2N_c\gg 1$ taken also to be large.  In AdS/CFT
\be
G_5=\frac{\pi L^3}{2N_c^2}.
\ee
The string length $l_s$  and the AdS radius $L$ are tied by $L^4/l_s^4=\lambda$ in the double
holographic limit. 
For a meaningful gauge theory result, $L$ which involves $l_s$ should drop at the end, and we will show this explicitly in our result shortly.

The 5d metric has the form
\be
ds^2={L^2\over z^2}\left({dz^2\over f}-fdt^2+d\vec x^2\right),
\ee
where the gauge theory spacetime is measured by $(t,\vec x)$. From this, one can obtain relations between the
local 5d energy $E_{AdS}$ and the 4d energy $E$ as follows

\be
E_{AdS}={z\over L} E.
\ee
Similarly, the 5d (proper) time and the 4d (proper) time are related as
\be
t_{AdS}={L\over z} t.
\ee
As a result, the 4-velocity $\dot{x}$ in the above result which is defined in terms of the 5d proper time
is roughly

\be
\dot{x}\approx  {z\over L}\gamma,
\ee
where $\gamma$ is the ordinary Lorenz factor in 4-dimensions. 
Also, since $m$ is the 5d rest energy it should be rescaled by $z/L$ to tie with the 4d rest energy $m_4$.
We now note that the square of the Riemann tensor in (\ref{SELF}) is of dimension but independent of $L$,

\be
{\bf R}^2 \approx \frac{{\bf P}(z/z_h)}{z^4},
\ee
with ${\bf P}$ a polynomial of $z/z_h\approx T/Q_s$. $Q_s$ is the jet saturation scale to be fixed  below.
The final integration $\int d\epsilon$ is of order $L/\gamma^2$ times some function of $T/Q_s$.
The $1/\gamma^2$ was explained before, whereas the factor $L$ follows the factor $L^2$ in $ds^2$,
since the geodesic motion is independent of the overall factor of $L^2$.

With these considerations, the 5-dimensional {\it non-covariant force} which  
is equal to 5-dimensional gravitational energy-loss takes the form

\be
{dE_{AdS}\over dt_{AdS}}&=& \left({z\over L}\right)^2{dE\over dt} \approx G_5 m^2 R^2 \left(\int d\epsilon\right)\left(\dot{x}\right)^4\nonumber\\
&\approx& {L^3\over N_c^2} \left(m_4 {z\over L}\right)^2 {1\over z^4} {L\over\gamma^2}\left({z\over L}\gamma\right)^4 {\bf F}(z/z_h)
\nonumber\\
&=& \left({z\over L}\right)^2{(m_4\gamma)^2\over N_c^2} {\bf F}(z/z_h).
\ee
With this, one can make a safe massless limit by replacing $m_4\gamma=E$ as it makes sense physically. Thus

\be
{dE\over dt} \approx \frac{E^2}{N_c^2} \,{\bf F}(z/z_h),\label{LONG}
\ee
with ${\bf F}(0)=0$. $L$ drops in the final result for the boundary observables. We set $L\equiv 1$ in the rest of our paper.
In thermal AdS$_5$ the universal function ${\bf F}$ appears
solely dependent on the ratio $z/z_h$ with the holographic direction playing the role
of a renormalization scale evolution. Indeed, along the longitudinal direction, the
jet forms at $z_i\approx 1/Q_s$ and ends at $z_h=1/(\pi T)$ the black-hole horizon.

Although (\ref{LONG}) is subleading in $1/N_c$, 
it is magnified by a large factor $\gamma^2\gg 1$ thereby dwarfing this suppression. Since the
gravitational radiation effect is classical, this longitudinal dragging force
is likely present at the boundary.  
We note that the work following from (\ref{LONG}) yields a power or intensity
\be
{\bf I}
\approx \frac{E^2}{N_c^2}\,\,\int_{z_i}^{z_h}\frac{dz}{\dot z}\,{\bf F}(z/z_h),
\label{XLONG}
\ee
in agreement with (\ref{COMP}). We cannot completly trace the origin of this agreement.

Finally, we note that the longitudinal {covariant} force following from  a dragging
colored string is of the order of $\gamma\sqrt{\lambda}T^2$~\cite{Gubser:2008as,Chesler:2008uy}. The ratio of the
longitudinal drag from gravitational radiation (or selfforce) to color is
\be
\frac{\rm gravity-radiation-drag}{\rm color-drag}\approx \frac{\gamma^2}{N_c^2\sqrt{\lambda}}\approx \frac{10^{2..4}}{10\times 5},
\ee
not small, for typical jets at RHIC and LHC with $\gamma=10-100$.  Of course,  the derivation 
given is perturbative, without back reaction explicitly included. (It means it
is only formally valid for $N_c,\lambda$ exceeding the realistic values of $N_c=3$ and $\lambda\approx 25$.) 

In summary, we argue that the gravitational radiation drag is comparable
to the color drag at RHIC, and is perhaps the dominant mechanism for 
jet loss at the LHC.

(Other effects subleading in $1/N_c$, such as the further terms supergravity Lagrangian beyond the Einstein-Hilbert Lagrangian, may also in principle lead to higher power of $\gamma$: we have not yet 
inverstigated their effect.)

\section{Matter versus wind frame for thermal AdS$_5$ }

We now proceed to analyze in detail the radiative gravitational force for thermal AdS$_5$ 
described by a black-brane with a horizon  located at some value  $z_h$ in the 5-th coordinate. 
Holography asserts that this classical set up is dual to thermal $\cal N$=4 supersymmeteric 
gauge theory at the boundary with temperature $T=1/(\pi z_h)$.

\subsection{The geodesics in the (plasma) matter rest frame}

To assess the effect of the thermal bath on the longitudinal induced gravitational self-force, 
we use in this section the standard thermal AdS$_5$ metric with plasma matter at rest. In terms of the 
coordinates $(t,x^1,x^2,x^3,z)$  the metric is
\be 
ds^2=\frac 1{z^2}\left(- fdt^2+(dx^1)^2 +(dx^2)^2+(dx^3)^2+\frac{dz^2}f \right),\nonumber\\
\ee
with $f=1-(z/z_h)^4$. 
In this frame the jet is represented by a point particle moving with some 4-velocity $\dot{x}^\mu=u^\mu$. At the start,  the jet moves  ultra-relativistically in the $x^1$ direction, so that $u^t,u^x=\gamma^1$ with $\gamma \gg 1$, while $u^z$ is small. As it proceeds, its longitudinal velocity decreases while its velocity along the holographic direction z direction increases, till finally it approaches the vicinity of the horizon. 

Using conservation of the 4-momentum of the particle one can readily derive the following expressions for the components of the 4-velocity 
\be \dot{t}&=& z^2 E /f, \,\,\,\dot{x}=z^2 P,\\
 \dot{z}^2&=& z^4 ( E^2- f P^2 ) -z^2 f, \ee
from which the path 
\be 
x^1=\int_{z_i}^z\,dz'\,\,\frac{P}{\sqrt{E^2-fP^2-f/{z'}^2}}. 
\ee
In pure AdS$_5$, $f=1$ leading to the Lorentz invariant combination
$E^2-P^2$ in the integrand. This is no longer the case in a thermal medium with $f\neq 1$.

An example of the corresponding trajectory in the x-z plane
is shown in Fig.\ref{fig_2geo} by the open circles. For comparison we also  show a geodesic for massless particles (blue asterisks).
Note that a jet here as a particle  starts from the initial value of the holographic coordinate $z_i$, taken to be 1, and ends at $z=z_h$.  In this example $z_h=5$. After selecting the jet gamma factor $E=40$ and fixing the initial $z_i=1$, the value of the jet momentum follows from the on-shell condition as  usual.
What we see  in Fig.\ref{fig_2geo} is qualitatively similar to the standard trajectory of a stone thrown from a cliff. The trajectory 
turns down as the increasing z-momentum takes over the conserved x-momentum.

The choice  of $z_i$ is a choice of the particle mass or jet scale (in the renormalization group sense) 
which is also referred to as the {\it saturation scale} $Q_s$.  Phenomenology
 sets it to be 1-2 GeV at RHIC and 2-3  GeV at LHC.  With the units thus fixed, the jet energy in this example should correspond to  40-80  GeV at RHIC and 80-120  GeV at  LHC. 
  The corresponding matter temperature is in this example $T=Q_s/5\pi$,  which is
 lower than the temperature occuring in real hadronic matter.
 We selected it, however, in order the
  stopping distance (about $24/Q_s$) be  (for $Q_s=1$ GeV )  about 5 fm, or
 comparable to the size of heavy nuclei used in the experiments.  
 
 The elapsed time (not shown) follows closely the distance 
except for a divergence as $z\rightarrow z_h$.  Any falling object freezes in the distant observer frame as it approaches the      horizon. In jet experiments, the time is usually limited by the matter freezeout time, thereby the falling jet remains close to $z_h$ but never crosses it.
  
The metric and thus all tensors are independent of gamma, and thus the counting of the power of gamma
is straightforward. It is reduced to counting the powers of the 4-velocities. For example 
\be  
\ddot{x}^a \approx \Gamma ^a_{bc} \dot{x}^b\dot{x}^c\approx \gamma^2 ,
\ee
so that $\ddot{x}\cdot\ddot{x}\approx \gamma^4$. Of course, one should check
that one gets nonzero result for an ultrarelativistic particle. An important exception 
involves contractions with the the Ricci tensor which has a special property $R_{ab}\approx  g_{ab}$.
Therefore its contraction ${\bf R}_{ab}\dot{x}^a\dot{x}^b\approx \gamma^0$ and not $\gamma^2$ as naively
expected.

\begin{figure}[!t]
\begin{center}
\includegraphics[width=7cm]{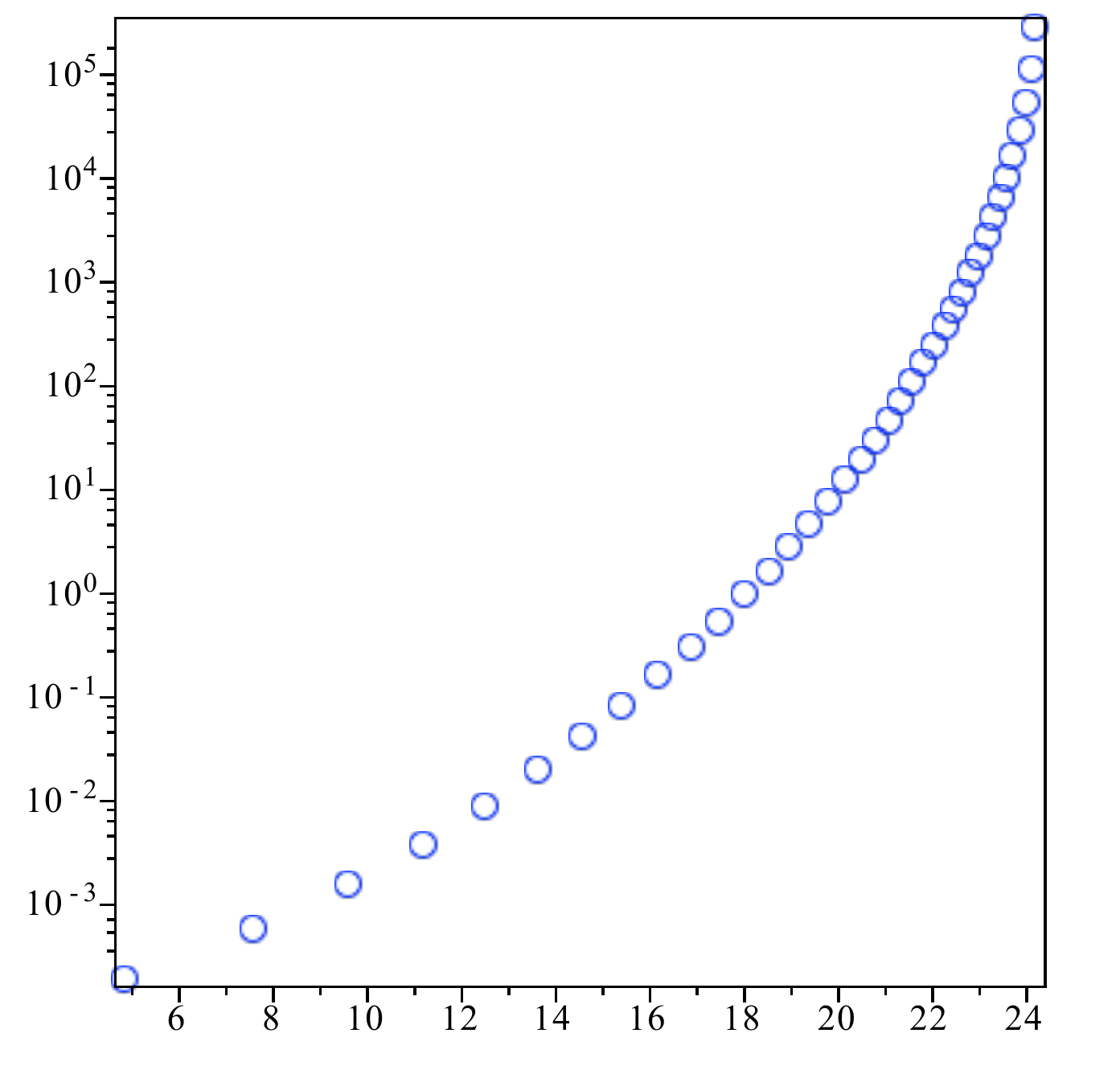}
\end{center}
\vspace{-5ex}\caption{
Dependence of the combination of two Riemann tensor and four velocities normalized to 
the leading power of the jet Lorentz factor $\gamma$, namely 
${\bf RR}uuuu/\gamma^4$, versus the longitudinal distance (in the $x^1$ direction)
travelled by the jet, for the numerical example displayed in Fig.1. The units, as  explained in the text, are fixed by the
particle mass or $Q_s$.} \label{fig_RRuuuu}
\end{figure}

Unlike the Ricci  terms, the one with the squared Riemann tensor does not have a simplified structure and thus retains its naive  order of magnitude $\approx \gamma^4$.
More specifically, we found
\ba  {\bf RR}uuuu=&&{\bf R}_{\sigma\gamma\kappa\delta}\,  g^{\sigma\nu} g^{\kappa\mu} \,{\bf R}_{\mu\alpha\nu\beta}  \, u^\alpha u^\beta u^\gamma u^\delta \\ \nonumber
= &&{\gamma^4{\bf P}\over  z^4(1-z^4/z_h^4)^2 },\ea
where ${\bf P}$ is some lengthy polynomial of degree 4 in the velocities with coefficients depending on $\gamma,z$. 
Instead of presenting it as such, we show in Fig~.\ref{fig_RRuuuu} its dependence on the longitudinal
distance $x^1$ travelled rather than $z$.  As we see from the plot, this contribution to the self-force not only depend strongly
 on the jet energy $\gamma$ but varies rapidly with the longitudinal distance crossed by the jet. It takes extremely large
 values near the jet final stopping point. Clearly, the back reaction from this force on the travelling jet needs to be taken into
 account.
 
\subsection{The ``hot wind" frame}

The dependence of all parameters in question on the jet energy (Lorentz factor $\gamma$) is also instructive to study in an
alternative frame, in which the longitudinal jet momentum $P_x$ can be set  to zero.  Such boosted frames were used before, e.g. in the study of moving charmonium \cite{Ejaz:2007hg}.

First, it is instructive to see how the cyclotron electromagnetic radiation is recovered in this frame. In the original frame all
powers of $\gamma$ follow from the velocity, while in the boosted frame the velocities are $\gamma^0$. The enhancement in this frame is transferred to the magnetic field which is magnified by $\gamma$.

The same happens with the boosted metric.  In empty AdS$_5$ the combination 
$-dt^2+d{x^1}^2$ turns to $-dt'^2+d{x^1}'^2$ after boosting, which is clearly Lorentz scalar.
In the boosted frame, the jet is at rest with all $\gamma$'s gone. Empty AdS$_5$ 
is Lorentz invariant with free particle motion being uneventful. 

At finite temperature, manifest Lorentz invariance in the metric is lost with
 \be 
 dt^2= (\gamma dt' -\sqrt{\gamma^2-1} d{x^1}')^2, 
 \label{METRIC}
 \ee 
and $d{x^1}^2=0$ after the boost. Thus the metric gets enhanced by $\gamma^2$ in this case. This
amounts to effectively boosting the energy density. In the rest frame of the jet the effective temperature is 
enhanced
\be 
T'=T \sqrt{\gamma}. 
\ee
Note that at large $\gamma$ the boosted metric in (\ref{METRIC}) has non-zero $g_{12}$ . There is "wind" in the x-direction.
This has obvious consequences on the particle motion. While it is still true that 
there is  conserved energy and momentum given by the lower components $u_0=E,u_1=P$, the upper components are mixed by the non-diaginal metric. As a result, even if $P=0$, one gets a nonzero velocity in the x direction since
\be   
u^0=g^{00} E,  \,\,\,  u^1=g^{10} E . 
\ee 
The upper component metric is obtained in the standard way with
\ba 
  u^t={dt \over d\tau}&&= E {z^2 (z_h^4+z^4\gamma^2-z^4) \over z^4-z_h^4} , \\ \nonumber 
  u^x= {dx \over d\tau}&&= E \sqrt{\gamma^2-1}{z^6 \over z^4-z_h^4}.                  
\ea
The geodesic path is again derived  from the on-shell condition $g_{ab} u^a u^b=-1$ 
after solving for the z-velocity
\ba (u^z)^2&&=({dz \over d\tau})^2 \\ \nonumber
&& =\left( -z^6 E^2+z^6 \gamma^2 E^2+z^4+E^2 z_h^4 z^2-z_h^4 \right){z^2\over z_h^4}. 
\ea
This is readily integrated. As usual, the ratio
$ (u^x/u^z)^2$ eliminates the proper time and yields the falling trajectory $x(z)$.

\begin{figure}[!t]
\begin{center}
\includegraphics[width=7cm]{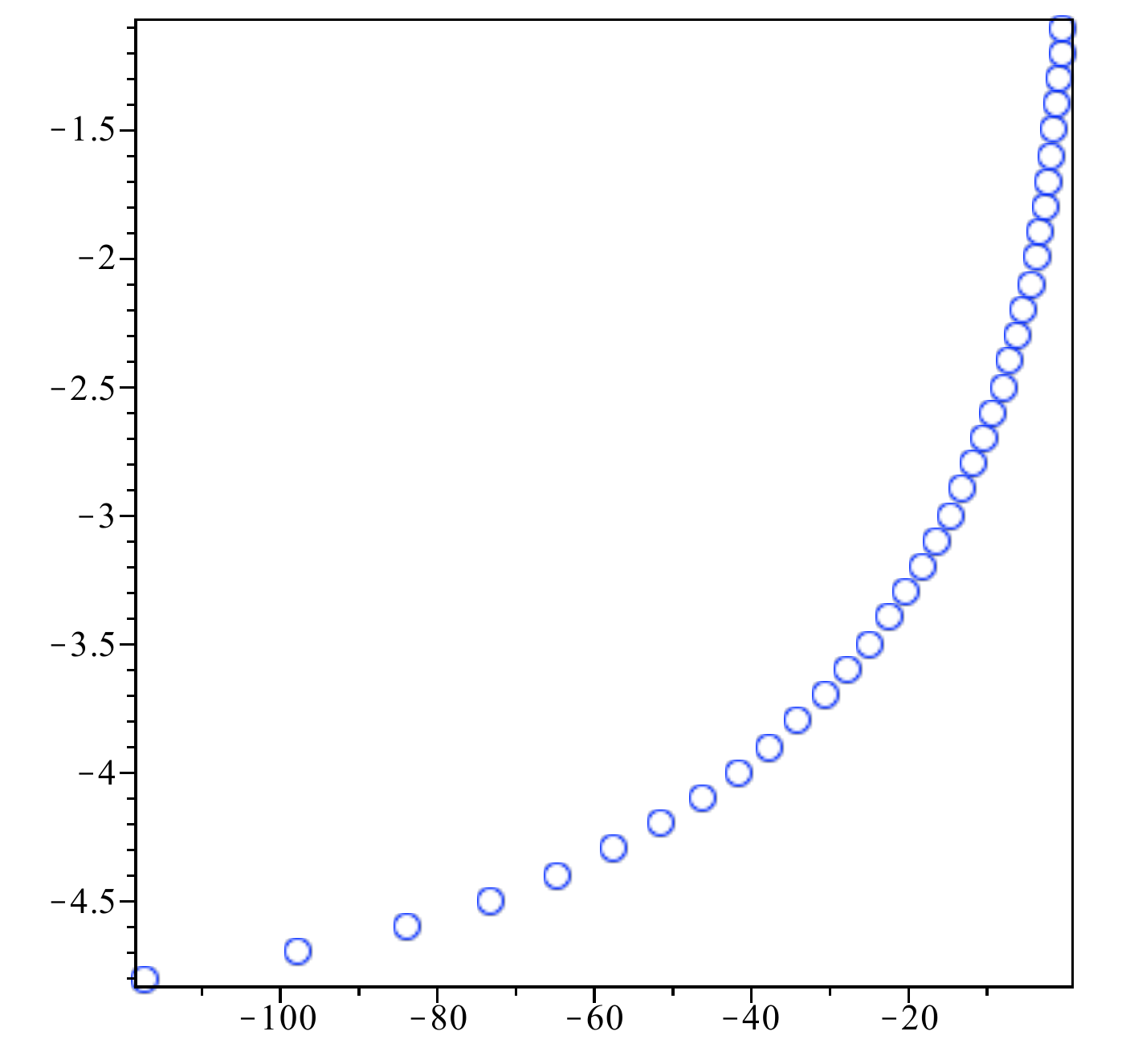}
\end{center}
\vspace{-5ex}\caption{(color online) The same example of a geodesic trajectory for a falling particle
as in the previous figure, but in the boosted ``hot wind" frame.    The points are in the (inverted) 5-th coordinate $-z$ as a function of $x^1$,
which is now also negative as the ``hot wind" blows in the negative  direction. 
}\label{fig_2geo_boosted}
\end{figure}

Fig.\ref{fig_2geo_boosted} shows the falling jet geodesic in the hot wind frame for the parameters used in the matter rest frame.
The motion is a free fall along the z-direction, followed by a horizontal motion along the wind direction. Initially, the particle is
nearly standing and non-relativistic with  $u^t \approx 1$ and all other components zero. Then these components get larger 
with typically $u^a\approx \gamma$.

In the initial time, all the $\gamma$ dependence is carried by the wind term in the metric 
 $\approx  (z/z_h)^4 \gamma^2$. Note that for our two Figures, the numerical value of the wind term is
 $\approx 40^2/5^4=2.56$ which is not very large compared to 1. It is 10 times more for twice more energetic jets.
we now detail how this combination is enetring the various contributions to the gravitational self-force
in this frame.

First, let's consider the particle acceleration along the geodesic
 \be  
 \ddot{x}^a \approx \Gamma^a_{bc} \dot{x}^b\dot{x}^c. 
 \ee
 Initially the velocities are of order $\gamma^0$, with all powers of $\gamma$ carried by the 
 Christoffels. Typically,
  \be 
  \Gamma^5_{11} ={(-z_h^4+z^4 )( z_h^4+z^4\gamma^2) \over z_h^8 z }.
  \ee
 Even when the particle is at rest with $\dot{x}^t=1$ (and all others zero) there is 
 an acceleration of order $\gamma^2$ along the z-direction.
 The calculation of the Riemann tensor contribution to the self-force 
 is straightforward in this frame. To leading order, the result is
\ba  
{\bf R}_{\sigma\gamma\kappa\delta}  g^{\sigma\nu} g^{\kappa\mu} {\bf R}_{\mu\alpha\nu\beta}   u^\alpha u^\beta u^\gamma u^\delta 
\approx \gamma^4 4 z^4 {\bf P'} ,
\ea
where ${\bf P'}$ is another fourth order polynomial in the velocities. For $z_h=1$ it is
\ba 
{\bf P'}=&&+3z^8(u^t)^4+3z^8(u^x)^4\nonumber \\ \nonumber
&&-12z^8(u^t)^3u^x+18z^8(u^x\,u^t)^2 \\ \nonumber
&&-12z^8(u^x)^3u^t-8z^4(u^x)^4 \\ \nonumber
&&-48z^4(u^x\,u^t)^2+32z^4(u^x)^3u^t \\ \nonumber
&&-8z^4(u^t)^4+32z^4(u^t)^3u^x \\ \nonumber
&&+42(u^x\,u^t)^2-28(u^t)^3u^x \\ \nonumber
&&+7(u^t)^4-28(u^x)^3u^t+7(u^x)^4. \\ 
\ea
It is non-zero even in the  initial part of the trajectory when the particle  is nearly
at rest $u^t=1,u^x=u^z=0$, as it has the $(u^t)^4$ term. However, as the wind accelerates the falling jet,
the polynomial is seen to keep the same power of $\gamma$. The quantity plotted
in Fig.2 is of course a scalar, so it is same in both frames.

%



\section{Summary and Discussion}
  Starting with a companion paper ~\cite{selfforce} we first mention its main result:
 the longitudinal selfforce in flat even-space+time
\be
&&{\rm 2+1:}\,\,\,\, (m\ddot{x}^a)_L\approx -\frac{2e^2}{\sqrt{3}}\frac{\ddot{x}\cdot\ddot{x}}{\sqrt{\ddot{x}\cdot\ddot{x}}}\,\dot{x}^a,\nonumber\\
&&{\rm 4+1:}\,\,\,\,  (m\ddot{x}^a)_L\approx -\frac{e^2}{10\sqrt{3}}\frac{\dddot{x}\cdot\dddot{x}}{\sqrt{\ddot{x}\cdot\ddot{x}}}\,\dot{x}^a,
\ee
for an electromagnetically charged particle undergoing circular motion in 2+1 and 4+1 dimensions respectively. Both of these
induced forces match the expected radiation asymptotically.  

Our main result is that the leading longitudinal gravitational selfforce (\ref{SELF}) with (\ref{COEF}), i.e.
\begin{eqnarray}
m\ddot{x}^a\approx-\frac{ G_5 m^2}{30\pi}\,\left(\frac 4{\sqrt{6\,\ddot{x}\cdot\ddot{x}}}\right)
\,{{{{\bf R}^m}_{e}}^{n}}_{b}{\bf R}_{mcnd}\,\,
\dot{x}^e\dot{x}^b\dot{x}^c\dot{x}^d\,\,\dot{x}^a,\nonumber\\
\end{eqnarray}
derived using similar arguments, starting from (\ref{XG1}).  The structure itself appears from
the  leading short distance singularity  of the propagator in 4+1 dimensions. Its part in brackets, coming from  resummation of the higher order terms, is schematic and we do not claim exactness of its
coefficient, just the power of $\gamma$. 
The convergence of the expansion follows from $\epsilon\approx 1/\gamma^2$ which is expected in the ultrarelativistic limit. Indeed, at large $\gamma$ the fields are Lorentz contracted causing them to be localized over a proper time of order $1/\gamma^2$.  In (\ref{COEF}) we have shown how this cutoff in $\epsilon$ is obtained through a qualitative resummation of the higher order terms. More work is required to fix this resummation quantitatively.

It is interesting to note the fundamentally different nature of the approximations leading to the 
radiation/self-force in the non-relativistic versus the relativistic limit. In the former, the radiation proceeds through
large wavelengths or small frequencies via a multipole expansion at large distances away from the charge/mass. 
In the latter, the expansion is dominated by the short wavelengths or frequencies, that is by the field in the
immediate vicinity of the charge/mass.  In a way, the two approaches are expected to be tied by the structures of energy-momentum conservation. 

To assess the magnitude of the gravitational self-force in AdS$_5$ we have used two different frames: the plasma matter rest
frame and the jet rest frame.  While scalars and vectors are expected to transform according to the expected lore of
relativity, the use of these two distinct frames shed light on the nature of the contributions to the self-force. The dominant
contribution to the covariant longitudinal gravitational drag is of order $\gamma^3$.  

The longitudinal self-force we have discussed is very different
from those discussed so far in the context of holography. In particular, it has nothing to do with the color charge of the jet but rather
with its energy. The massive amount of excitation left behind by the jet produces backward
gravitational drag or pull, which we have found to be growing dramatically with the jet  energy. This effect is classical,
large and physically well separated from the variety of other subleading effects one may think of. It is by no means 
subleading although in holography it is formally down by a power of Newton's gravitational constant $G_5\approx 1/N_c^2$.

The present work is the first indication for the existence of strong self-induced gravitational forces in thermal
AdS$_5$ that have a power of the Lorentz factor $\gamma$
larger than one. Many of the terms with higher 
powers do appear for arbitrary paths but are canceled out in the result, mostly because
on geodesics there is no covariant acceleration. The more involved problem of a dropping string with a 
longitudinal color drag is likely to depart from the geodesic motion and therefore involves higher powers. This issue
will be addressed elsewhere.

It is amusing to note that the $\gamma^2$ dependence of the non-covariant force and hence the intensity
radiated by energy conservation is the same as in 3+1 Schwartzschield case~\cite{Khriplovich:1973qf}. We
cannot trace the connection except the fact that the combination $(\gamma\,m)^2=E^2$ in the radiated intensity is
the only one that is finite in the massless limit $m\rightarrow 0$ and $\gamma\rightarrow \infty$ with $E$ fixed.

This paper is intended as a qualitative indication for the existence and importance of classical gravitational self-forces
in studying jet quenching via AdS/CFT correspondence.
We have not tried to assess their effects on jets trajectories in the bulk or their stopping distance on the boundary quantitatively. For realistic values of $N_c$ and $\lambda$, 
(\ref{SELF}) appears to be somewhat too strong for ultrarelativistic jets to be treated in perturbation theory. Therefore more work is needed to assertain the role of these forces in a reliable calculational framework.
  
 It is very challenging to try to understand the origin of this effect from
the gauge theory side without recourse to holography and bulk gravity.
As a parting comment, let us speculate that bulk gravitons and gravity are
perhaps related to Pomerons and effective Pomeron field theory, which 
practitioners in the field have tried to construct for the past many decades.

\vskip 1cm {\bf Acknowledgments} \vskip .2cm 
We thank Gokce Basar, Dima Kharzeev, Derek Teaney, and Raju Venugopalan for discussions.
This work was
supported in parts by the US-DOE grant DE-FG-88ER40388.

\vskip 1cm

\section{Appendix }

The structure of the singular terms in even-space+time dimensions can be readily obtained using
the same arguments as in odd-space+time dimensions. Indeed, the retarded propagator in arbitrary
space-time dimensions follows from 
\be
G^-(x,x')&&=-\,\Theta(x,x')\,{\rm Im}\,G_F(x,x')\nonumber\\
&&=-\,\Theta(x,x')\,{\rm Im}\,G_H(x,x')|_{\sigma+i0},
\label{Ap1}
\ee
following DeWitt's prescription between the Feynman and Hadamard propagator.  
In general AdS$_5$ the singular part of the Hadamard propagator reads
\be
G_H(x,x')=\frac{g(\sigma)}{(-2\sigma)^{3/2}}+w(\sigma),
\label{Ap2}
\ee
where $\sigma$ is the world function bi-scalar (\ref{G5}). Both $g(\sigma)$ and $w(\sigma)$ are regular
functions of $\sigma$.   $g(\sigma)$ is fixed by $\Box G_H=0$ in leading order 
\be
2\,\partial^a g\,\sigma_a +g(\Box \sigma-5)=0,
\label{Ap3}
\ee
after using that $\sigma^a\sigma_a=2\sigma$. Since
\be
\Box\sigma=5-\sigma^a\partial_a\,{\rm ln}\Delta,
\label{Ap4}
\ee
it follows that 
\be
\partial_a\,{\rm ln}\,(g/{\sqrt{\Delta}})=0,
\label{Ap5}
\ee
which fixes $g$ up to an overall normalization.


\end{document}